\documentclass[astronomy,article,accept,pdftex,moreauthors,astronomy]{Definitions/mdpi}
\firstpage{1}
\pubvolume{1}
\issuenum{1}
\articlenumber{0}
\pubyear{2023}
\copyrightyear{2023}
\externaleditor{Academic Editor: Ignatios Antoniadis}
\datereceived{27 July 2022}
\daterevised{27 September 2022}
\dateaccepted{}
\datepublished{}
\hreflink{https://doi.org/} % If needed use \linebreak
\Title{Isoscalar Giant Monopole Resonance in Spherical Nuclei as a
Nuclear Matter Incompressibility Indicator}

\TitleCitation{Isoscalar Giant Monopole Resonance in Spherical Nuclei as a
Nuclear Matter Incompressibility Indicator}

\Author{Mitko K. Gaidarov %MDPI: 1. Please carefully check the accuracy of names and affiliations. 2. we suggest to use the full name word to replace the abbreviation names, please confirm and revise.
 *\orcidB{}, Martin V. Ivanov\orcidC{}, Yordan I. Katsarov and Anton N. Antonov\orcidD{}}

\AuthorNames{M.K. Gaidarov, M.V. Ivanov, Y.I. Katsarov and A.N. Antonov}

\AuthorCitation{Gaidarov, M.K.; Ivanov, M.V.; Katsarov, Y.I.; Antonov, A.N.}
\address[1] {%
Institute for Nuclear Research and Nuclear Energy, Bulgarian Academy of Sciences, 1784 Sofia, Bulgaria; martin.inrne@gmail.com (M.V.I.); yordankatsarov@gmail.com (Y.I.K.); antonovshumen@gmail.com (A.N.A.)}

\corres{\hangafter=1 \hangindent=1.05em \hspace{-0.82em}Correspondence: gaidarov@inrne.bas.bg}

\abstract{The incompressibility of both nuclear matter and finite
nuclei is estimated by the monopole compression modes in nuclei in
the framework of a nonrelativistic Hartree--Fock--Bogoliyubov
method and the coherent density fluctuation model. The monopole
states originate from vibrations of the nuclear density. The
calculations in the model for the incompressibility in finite
nuclei are based on the Brueckner energy--density functional for
nuclear matter. Results for the energies of the breathing
vibrational states and finite nuclei incompressibilities are
obtained for various nuclei and their values are compared with
recent experimental data. The evolution of the isoscalar giant
monopole resonance (ISGMR) along Ni, Sn, and Pb isotopic chains is
discussed. This approach can be applied to analyses of neutron
stars properties, such as incompressibility, symmetry energy,
slope parameter, and other astrophysical quantities, as well as
for modelling dynamical behaviors within stellar environments.}


\begin{document}
%
%%%%%%%%%%%%%%%%%%%%%%%%%%%%%%%%%%%%%%%%%%%

\section{Introduction}

In recent years, experimental and theoretical studies of giant
resonances have become a rich source of information on the
collective response of the nucleus to its density \linebreak fluctuations
\cite{Bohr75,Harakeh2001}. In particular, the isoscalar giant
monopole resonance (ISGMR) plays an important role in constraining
the nuclear equation of state (EOS)
\cite{Harakeh2001,Brandenburg82,Zwarts83,Brandenburg83,Shlomo93,Youngblood99}.
An important issue is that the energy of this resonance is closely
related to the nuclear incompressibility. The latter can be
connected to the incompressibility of the infinite nuclear matter,
which represents an important ingredient of the nuclear matter
\textls[-5]{EOS. It is well known that the EOS plays a crucial role in the
description of astrophysical quantities, such as radii and masses
of neutron stars, the collapse of the heavy stars in super novae
explosions, as well as in modeling of heavy-ion collision. The
20\% uncertainty of the currently accepted value of the
incompressibility of nuclear matter is largely driven by the poor
determination of the EOS isospin asymmetry term. Therefore, to
make this term more precise, recent experimental measurements of
isoscalar monopole modes are being extended in isotopic chains
from the nuclei on the valley of stability towards exotic nuclei
with larger proton--neutron asymmetry.}

The isoscalar resonances are excited through low-momentum transfer
reactions in inverse kinematics, that require special detection
devices. At present, promising results have been obtained using
active targets. Different measurements have been conducted on Ni
isotopes far from stability, namely $^{56}$Ni
\cite{Monrozeau2008,Bagchi2015} and $^{68}$Ni
\cite{Vandebrouck2014,Vandebrouck2015}. In particular, the
$^{68}$Ni experiment is the first measurement of the isoscalar
monopole response in a short-lived neutron-rich nucleus using
inelastic alpha-particle scattering. The peak of the ISGMR was
found to be fragmented, indicating a possibility for a soft
monopole resonance.

The discussion on how to extract the incompressibility of nuclear
matter $\Delta K^{NM}$ from the ISGMR dates back to the years
1980s \cite{Blaizot80} (see also more recent review
\cite{Garg2018}). The measurement of the centroid energy of the
ISGMR
\cite{Li2010,Patel2012,Blaizot76,Button2017,Howard2019,Howard2020a,Howard2020b}
provides %MDPI: This is an important note. The references numbers should appear in numerical order. Reference 16 is missing, please confirm and add.
a very sensitive method to determine the value of $\Delta
K^{NM}$. Theoretical investigations in various \linebreak models
\cite{Brueckner70,Shlomo2001,Chen2012,Anders2013,Su2018,Colo2020,Bonasera2021}
with grouped values of the nuclear matter incompressibility
$\Delta K^{NM}$ predict different ISGMR energies. In comparison
with the experimental data, one could give the constraint on the
nuclear matter incompressibility.

In the present work, the incompressibility and the centroid energy
of ISGMR are investigated for three isotopic chains on the basis
of the Brueckner energy-density functional for nuclear matter
\cite{Brueckner68,Brueckner69} and using the coherent density
fluctuation model (CDFM) \linebreak (e.g., Refs.~\cite{Ant79,Ant80,Ant82,Ant85,Ant89,Ant94,AHP1,AHP2}). This
method is a natural extension of the Fermi gas model based on the
delta-function limit of the generator coordinate method
\cite{AHP1,AHP2,Grif57} and includes long-range correlations of
collective type. During the years the CDFM has been successfully
applied to calculations of nuclear structure and nuclear reactions
characteristics. Among them we would like to note the calculated
energies, density distributions and rms radii of the ground state in
$^{4}$He, $^{16}$O, and $^{40}$Ca nuclei \cite{Ant88}. Here, we mention
particularly the calculations within the CDFM of the energies of breathing
monopole states in $^{16}$O, $^{40}$Ca, $^{90}$Zr, $^{116}$Sn, and $^{208}$Pb
performed in Ref.~\cite{Ant91} and presented also in Chapter 8 of Ref.~\cite{AHP2}.
In the latter are also given references for experimental data and other theoretical results
available until the early 1990s. Concerning the reaction properties, the CDFM has
been employed in Refs.~\cite{Ant2004,Ant2005} to calculate the
scaling function in nuclei using the relativistic Fermi gas
scaling function, which has been applied to lepton scattering
processes
\cite{Ant2004,Ant2005,Ant2006a,Ant2006b,Ant2008,Ant2007,Ant2009}.
In addition, information about the role of the nucleon momentum
and density distributions for the explanation of superscaling in
lepton--nucleus scattering has been obtained
\cite{Ant2005,Ant2006a}, also in studies of cross sections for
several reactions: inclusive electron scattering in the
quasielastic and $\Delta$ regions \cite{Ant2006b,Ant2008} and
neutrino (antineutrino) scattering both for charge-changing
\cite{Ant2008,Ant2009} and for neutral-current
\cite{Ant2007,Ant2009} processes. Furthermore, the CDFM was applied to
study the scaling function and its connection with the spectral
function and the nucleon momentum \linebreak distribution \cite{Cab2010}.

The efficiency of CDFM to be applied as a ``bridge'' for a
transition from the properties of nuclear matter to the properties
of finite nuclei studying the nuclear symmetry energy (NSE), the
neutron pressure, and the asymmetric compressibility in finite
nuclei was demonstrated in our previous works
\cite{Gaidarov2011,Gaidarov2012,Gaidarov2014,Antonov2016,Antonov2018,Danchev2020,Gaidarov2020,Gaidarov2020ch}.
Although there is enough collected information for the mentioned
EOS quantities, the volume and surface symmetry energies have been
poorly investigated till now. In Ref.~\cite{Gaidarov2021} we
proposed a new alternative approach to calculate the ratio of the
surface to volume components of the NSE in the framework of the
CDFM. We have demonstrated that the new scheme provides more
realistic values, in a better agreement with the empirical data,
and exhibits correct conceptual advantages.

In this work, we perform calculations and give results for the
excitation energies of ISGMR for Ni, Sn, and Pb isotopes. Our main
task is to validate the CDFM for studies of collective vibrational
modes by using as a main theoretical ground the self-consistent
Hartree--Fock (HF)+BCS method with Skyrme interactions. The
mentioned above model gives a link between nuclear matter and
finite nuclei in studying of their properties, such as binding
energies and rms radii of light, medium, and heavy nuclei. As an
example, for nuclear matter we adopt the energy-density functional
(EDF) of Brueckner {et al.} \cite{Brueckner68,Brueckner69}.
Obviously, more realistic functionals should be employed in the
future studies which would lead to values of the excitation
energies of ISGMR that are in better agreement with the experimental
ones. More details on this point are given in the last section of
the work, where specific future improvements are pointed out. We present
and discuss the values of the centroid energies in Sn isotopic
chain ($A$=112-124) studying its isotopic sensitivity. The main
reason to select these chains of spherical nuclei is partly
supported by their recent intensive ISGMR measurements so that we
focus too on the comparison with the available experimental data
for Ni \cite{Lui2006}, Sn \cite{Cao2012}, and Pb
\cite{Fujiwara2011,Youngblood2004} isotopes.

In the next Section~\ref{sec:2} we give definitions of the excitation energy
of ISGMR and EOS parameters of nuclear matter that characterize
its density dependence around normal nuclear matter density, as
well as a brief description of the CDFM formalism that provides a
way to calculate the finite nuclei quantities. The numerical
results are presented and discussed in Section~\ref{sec:3}. The main
conclusions of the study are summarized in Section~\ref{sec:4}.

\section{Theoretical Formalism}\label{sec:2}

\subsection{Excitation Energy of the ISGMR}\label{sec:2.1}

The centroid energy of ISGMR, $E_{ISGMR}$ is generally related to
a finite nucleus incompressibility $\Delta K(N,Z)$ for a nucleus
with $Z$ protons and $N$ neutrons ($A=Z+N$ is the mass number).
Among the various definitions of $E_{ISGMR}$ we will mention the
one from, e.g., Ref.~\cite{Brueckner70}):
\begin{equation}
E_{ISGMR}=\frac{\hbar}{r_{0}A^{1/3}}\sqrt{\frac{\Delta
K(N,Z)}{m}},
\label{eq:1b}
\end{equation}
where $r_{0}$ is deduced from the equilibrium density and $m$ is
the nucleon mass. The excitation energy of the ISGMR is also
expressed in the scaling model \cite{Stringari82} as (in
Refs.~\cite{Patel2012,Blaizot76}, for instance)
\begin{equation}
E_{ISGMR}=\hbar \sqrt{\frac{\Delta K(N,Z)}{m<r^{2}>}},
\label{eq:1a}
\end{equation}
where $<r^{2}>$ denotes the mean square mass radius of the nucleus
in the ground state. Depending on the adopted model, the value of
$E_{ISGMR}$ is associated with different moment ratios of the
ISGMR strength distribution. Its extraction is the main focus of
the experiments, which aim to constrain the incompressibility of
the infinite nuclear matter and, as a consequence, the EOS
\cite{Garg2018}. Particularly, it should be noticed that
definition (\ref{eq:1a}) is usable under the assumption that the
strength distribution of a given multipolarity of the resonance is
contained within a single collective peak \cite{Howard2019}.

\subsection{The Key EOS Parameters in Nuclear Matter}\label{sec:2.2}

The symmetry energy $S(\rho)$ is defined by the energy per
particle for nuclear matter (NM) $E(\rho,\delta)$ in terms of the
isospin asymmetry $\delta=(\rho_{n}-\rho_{p})/\rho$
\begin{equation}
S(\rho)=\frac{1}{2}\left.
\frac{\partial^{2}E(\rho,\delta)}{\partial\delta^{2}} \right
|_{\delta=0}, \label{eq:3}
\end{equation}
where
\begin{equation}
E(\rho,\delta)=E(\rho,0)+S(\rho)\delta^2+O(\delta^4)+
\cdot\cdot\cdot \label{eq:4}
\end{equation}
and $\rho=\rho_{n}+\rho_{p}$ is the baryon density with $\rho_{n}$
and $\rho_{p}$ denoting the neutron and proton densities,
respectively (see, e.g., \cite{Gaidarov2021,Diep2003,Chen2011}).

The incompressibility (the curvature) of the symmetry energy
$\Delta K^{NM}$ is given by
\begin{equation}
\Delta K^{NM}=9\rho_{0}^{2}\left.
\frac{\partial^{2}S}{\partial\rho^{2}} \right |_{\rho=\rho_{0}},
\label{eq:5}
\end{equation}
where $\rho_{0}$ is the density at equilibrium.

\subsection{The EOS Parameters of Finite Nuclei in the Coherent Density Fluctuation Model}\label{sec:2.3}

The CDFM was suggested and developed in Refs.~\cite{Ant79,Ant80,Ant82,Ant85,Ant89,Ant94,AHP1,AHP2}
(see also our recent \linebreak papers
\cite{Gaidarov2012,Danchev2020,Gaidarov2021}). In it the one-body
density matrix (OBDM) of the nucleus $\rho({\bf r},{\bf
r^{\prime}})$
\begin{equation}
\rho({\bf r},{\bf r^{\prime}})=\int_{0}^{\infty}dx |F(x)|^{2}
\rho_{x}({\bf r},{\bf r^{\prime}})
\label{eq:6}
\end{equation}
is expressed by OBDM's of spherical ``pieces'' of nuclear matter
(``fluctons'') with radius $x$ of all $A$ nucleons uniformly
distributed in it:
\begin{equation}
\rho_{x}({\bf r},{\bf r^{\prime}})=3\rho_{0}(x)
\frac{j_{1}(k_{F}(x)|{\bf r}-{\bf r^{\prime}}|)}{(k_{F}(x)|{\bf
r}-{\bf r^{\prime}}|)} \Theta \left (x-\frac{|{\bf r}+{\bf
r^{\prime}}|}{2}\right ).
\label{eq:7}
\end{equation}

In Equation~(\ref{eq:7}) $j_{1}$ is the first-order spherical Bessel
function and
\begin{equation}
k_{F}(x)=\left(\frac{3\pi^{2}}{2}\rho_{0}(x)\right )^{1/3}\equiv
\frac{\alpha}{x}
\label{eq:8}
\end{equation}
is the Fermi momentum with\vspace{-6pt}
\begin{equation}
\alpha \equiv \left(\frac{9\pi A}{8}\right )^{1/3}\simeq 1.52A^{1/3}.
\label{eq:9}
\end{equation}

It can be seen from Equation~(\ref{eq:6}) that the density distribution
in the CDFM is:
\begin{equation}
\rho({\bf r})=\int_{0}^{\infty}dx|F(x)|^{2}\rho_{0}(x)\Theta
(x-|{\bf r}|)
\label{eq:10}
\end{equation}
with\vspace{-6pt}
\begin{equation}
\rho_{0}(x)=\frac{3A}{4\pi x^{3}}.
\label{eq:11}
\end{equation}

It follows from Equation~(\ref{eq:10}) that the weight function
$|F(x)|^{2}$ of CDFM can be obtained in the case of monotonically
decreasing local densities ({\it i.e.}, for $d\rho(r)/dr\leq 0$)
by
\begin{equation}
|F(x)|^{2}=-\frac{1}{\rho_{0}(x)} \left. \frac{d\rho(r)}{dr}\right
|_{r=x}
\label{eq:12}
\end{equation}
being normalized as
\begin{equation}
\int_{0}^{\infty}dx |F(x)|^{2}=1.
\label{eq:13}
\end{equation}

In the case of the Brueckner method for nuclear matter energy
\cite{Brueckner70,Brueckner68,Brueckner69} the symmetry energy
$S^{NM}(x)$ of NM with density $\rho_{0}(x)$ is (see, e.g.,
Refs.~\cite{Gaidarov2011,Danchev2020}):
\begin{equation}
S^{NM}(x)=41.7\rho_{0}^{2/3}(x)+b_{4}\rho_{0}(x)+b_{5}\rho_{0}^{4/3}(x)+b_{6}\rho_{0}^{5/3}(x).
\label{eq:14}
\end{equation}

Then, correspondingly, the asymmetric incompressibility has the
form \cite{Gaidarov2011,Gaidarov2012}:
\begin{equation}
\Delta
K^{NM}(x)=-83.4\rho_{0}^{2/3}(x)+4b_{5}\rho_{0}^{4/3}(x)+10b_{6}\rho_{0}^{5/3}(x).
\label{eq:15}
\end{equation}

The expression for the energy density of the method of Brueckner
\cite{Brueckner68,Brueckner69} (see \linebreak also
\cite{Gaidarov2011,Gaidarov2012,Bethe71}), which is used to obtain
Equations (\ref{eq:14}) and (\ref{eq:15}) from Equations (\ref{eq:3}) and
(\ref{eq:5}), correspondingly, contains the following values of
the parameters:
\vspace{-6pt}
\begin{eqnarray}
b_{1}&=&-741.28, \;\;\; b_{2}=1179.89, \;\;\; b_{3}=-467.54,\nonumber \\
b_{4}&=&148.26, \;\;\;\;\;\; b_{5}=372.84, \;\;\;\; b_{6}=-769.57.
\label{eq:16}
\end{eqnarray}

According to the CDFM scheme, the symmetry energy and the
curvature for finite nuclei can be expressed in the following
forms:
\begin{equation}
s=\int_{0}^{\infty}dx|F(x)|^{2}S^{NM}(x),
\label{eq:17}
\end{equation}
\begin{equation}
\Delta K=\int_{0}^{\infty}dx|F(x)|^{2}\Delta K^{NM}(x).
\label{eq:18}
\end{equation}

In our calculations we apply self-consistent deformed Hartree--Fock
method with density-dependent Skyrme interactions \cite{vautherin}
with pairing correlations. We use the Skyrme SLy4 \cite{sly4}, Sk3
\cite{sk3} and SGII \cite{sg2} parametrizations (see also
\cite{Gaidarov2011,Gaidarov2012,Gaidarov2014,Antonov2016,Danchev2020,Sarriguren2007}).
In addition, we probe the SkM parameter set \cite{Krivine80},
which led to an appropriate description of bulk nuclear
properties. All necessary expressions for the single-particle
functions and densities in the HF+BCS method can be found, e.g.,
in Ref.~\cite{Gaidarov2011}.

It is known that the value of the nuclear matter incompressibility
$\Delta K^{NM}$ plays a key role in determining the location of
the ISGMR centroid energy \cite{Cao2012}. The different Skyrme
parameter sets used in the present calculations are chosen since
they are characterized by different values of the nuclear
incompressibility, $\Delta K^{NM}$ = 230, 217, 215, and 355 MeV for
SLy4, SkM, SGII, and Sk3, respectively, \cite{Danielewicz2009}.

The mean square radii for protons and neutrons are defined as\vspace{-6pt}
\begin{equation}
<r_{\rm p,n}^2> =\frac{ \int R^2\rho_{\rm p,n}({\vec R})d{\vec R}}
{\int \rho_{\rm p,n}({\vec R})d{\vec R}}.
\label{eq:19}
\end{equation}
The matter mean square radius $<r^{2}>$ entering Equation~(\ref{eq:1a})
can be calculated by\vspace{-6pt}
\begin{equation}
<r^{2}>=\frac{N}{A}<r_{n}^{2}>+\frac{Z}{A}<r_{p}^{2}>.
\label{eq:20}
\end{equation}

As shown in Section \ref{sec:2.1}, there exist two ways to calculate the
excitation energy of the giant monopole resonance. In both
definitions the finite nuclei incompressibility $\Delta K$
(Equation~(\ref{eq:18})) is obtained within the CDFM. In the present
work, describing the monopole vibrations in terms of harmonic
oscillations of the nuclear size and assuming an $A^{1/3}$ law for
it, we calculate $E_{ISGMR}$ by using Equation~(\ref{eq:1b}). In it
values of the parameter $r_{0}$ between 1.07 and 1.2 fm are adopted,
which are determined from experiments on particle scattering off nuclei.
If one applies definition (\ref{eq:1a}), then the mean square mass radius
(Equation~(\ref{eq:20})) has to be used.

%%%%%%%%%%%%%%%%%%%%%%%%%%%%%%%%%%%%%%%%%%
\section{Results and Discussion}\label{sec:3}

Here we present the obtained results for the centroid energies of
the ISGMR in finite nuclei extracted from nuclear matter many-body
calculations using the Brueckner EDF. We show also their isotopic
sensitivity for Ni, Sn, and Pb chains.

\begin{figure}[H]
\begin{adjustwidth}{-\extralength}{0cm}
\centering
\includegraphics[width=1.25\textwidth]{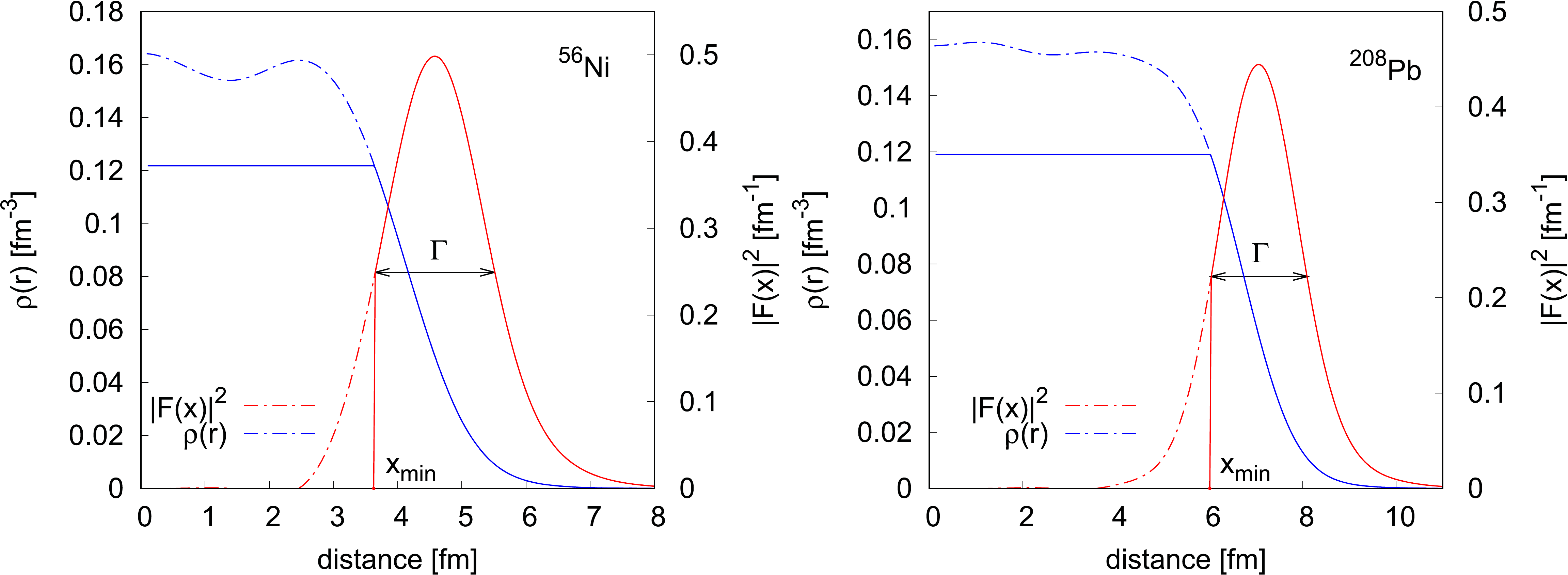}
\end{adjustwidth}
\caption{The densities $\rho(r)$ (in fm$^{-3}$) of $^{56}$Ni and
$^{208}$Pb calculated in the Skyrme HF + BCS method with SLy4
force (normalized to $A=56$ and $A=208$, respectively) and the
weight function $|F(x)|^{2}$ (in fm$^{-1}$) normalized to unity
(Equation~(\ref{eq:13})).\label{fig1}}
\end{figure}

First, in Figure~\ref{fig1} we overlay, as examples, the density
distributions of $^{56}$Ni and $^{208}$Pb and the corresponding
CDFM weight function $|F(x)|^{2}$ as a function of $x$. As
mentioned before, the densities are obtained in a self-consistent
Hartree--Fock+BCS calculations with SLy4 interaction. The function
$|F(x)|^{2}$ which is used in Equation~(\ref{eq:18}) to obtain the
incompressibility modulus, which is necessary to calculate the
$E_{ISGMR}$, has the form of a bell with a maximum around
$x=R_{1/2}$ at which the value of the density $\rho(x=R_{1/2})$ is
around half of the value of the central density equal to
$\rho_{c}$ $[\rho(R_{1/2})/\rho_{c}=0.5]$. It was shown in
Refs.~\cite{Danchev2020,Gaidarov2021} that in this region around
$\rho=\rho_{c}/2$ the values of $\Delta K^{NM}(\rho)$ take a
significant part in the calculations. This fact is of particular
importance and is related to the behavior of $S^{NM}(x)$
(Equation~(\ref{eq:14})) in the case of the Brueckner EDF showing its
isospin instability (see Figure~1 of Ref.~\cite{Gaidarov2021}), in
contrast with other more realistic energy-density functionals.
Therefore, to fully specify the role of both quantities $\Delta
K^{NM}[\rho_{0}(x)]$ and $|F(x)|^{2}$ in the expression
(\ref{eq:18}) for the finite nuclei incompressibility $\Delta K$
and to locate the relevant region of densities in finite nucleus
calculations, we apply the same physical criterion related to the
weight function $|F(x)|^{2}$, as in \cite{Gaidarov2021}. This is
the width $\Gamma$ of the weight function $|F(x)|^{2}$ at its half
maximum (which is illustrated in Figure~\ref{fig1} on the example of
$^{56}$Ni and $^{208}$Pb nuclei together with the corresponding
distance in the density distribution $\rho(r)$), which is a good
and acceptable choice. More specifically, we define the lower
limit of integration as the lower value of the radius $x$,
$x_{min}$, corresponding to the left point of the half-width
$\Gamma$ (for more details see the discussion in
Refs.~\cite{Danchev2020,Gaidarov2021}). One can see also in
Figure~\ref{fig1} the part of the density distribution $\rho(r)$ (at
$r\geq x_{min}$) that is involved in the calculations.

The centroid positions of the monopole mode obtained in this work
are compared with available experimental data in
Tables~\ref{tab1}--\ref{tab3}. The calculated values of
$E_{ISGMR}$ with SLy4 and SkM forces for Ni and Pb isotopes are
given in Tables~\ref{tab1} and \ref{tab3}, respectively. The
values of the centroid energies for Sn isotopes obtained from
calculations with three Skyrme interactions (SLy4, SGII, Sk3) are
listed in Table~\ref{tab2}. It can be seen from Table~\ref{tab1}
that a very good agreement with the experimental data for
$^{56,58,60}$Ni is obtained, while the results with both Skyrme
interactions underestimate the experimental energy of the soft
monopole vibrations of $^{68}$Ni. The excitation energy of this
ISGMR in $^{68}$Ni is located unexpectedly at higher energy \linebreak (21.1
MeV) for the Ni isotopic chain, having at the same time large
error bars. The reason is due to the large fragmentation of the
isoscalar monopole strength in the unstable neutron-rich $^{68}$Ni
nucleus, much more than in stable nuclei
\cite{Vandebrouck2014,Vandebrouck2015}. The obtained values of
$E_{ISGMR}$ for Sn isotopes ($A$ = 112--124) exhibit small difference
regarding the Skyrme parametrization (see Table~\ref{tab2}). The
theoretical results for the centroid energies for the same Sn
isotopes obtained in Ref.~\cite{Cao2012} by using the SkP (between
14.87 and 15.60 MeV), SkM* (between 15.57 and 16.23 MeV), and SLy5
(between 15.95 and 16.61 MeV) parameter sets are in good agreement
with our results. Almost no dependence on the Skyrme forces used
in the calculations of the centroid energies is found for Ni and
Pb isotopes being slightly larger in the case of SkM interaction
than when using the SLy4 one.

\begin{table}[H]
\caption{The values of the centroid energies $E_{ISGMR}$ (in MeV)
of Ni isotopes obtained from HF+CDFM calculations in this work
using SLy4 and SkM Skyrme forces compared with the experimental
data found in the literature.\label{tab1}}
\newcolumntype{C}{>{\centering\arraybackslash}X}
\begin{tabularx}{\textwidth}{CCCC}
\toprule
\textbf{Nucleus} & \textbf{SLy4}  & \textbf{SkM}  & \textbf{Exp.}  \\
\midrule
$^{56}$Ni        & 19.41      & 19.57   & 19.1 $\pm $ 0.5 \cite{Bagchi2015}                       \\
                 &            &         & 19.3 $\pm $ 0.5 \cite{Monrozeau2008}                    \\
$^{58}$Ni        & 18.95      & 19.18   & 18.43 $\pm $ 0.15 \cite{Lui2006}                        \\
$^{60}$Ni        & 18.62      & 18.79   & 18.10(29)  \cite{Lui2006}                               \\
$^{68}$Ni        & 17.46      & 17.70   & 21.1 $\pm $ 1.9 \cite{Vandebrouck2014,Vandebrouck2015}  \\
\bottomrule
\end{tabularx}
\end{table}
\vspace{-10pt}

\begin{table}[H]
\caption{The values of the centroid energies $E_{ISGMR}$ (in MeV)
of Sn isotopes ($A$=112-124) obtained from HF+CDFM calculations in
this work using SLy4, SGII, and Sk3 Skyrme forces. The
experimental data are taken from Table III of
Ref.~\cite{Cao2012}.\label{tab2}}
\newcolumntype{C}{>{\centering\arraybackslash}X}
\begin{tabularx}{\textwidth}{CCCCC}
\toprule
\textbf{Nucleus} & \textbf{SLy4}  & \textbf{SGII}   & \textbf{Sk3}   & \textbf{Exp.}  \\
\midrule
$^{112}$Sn    & 15.04      & 15.30   & 14.89     & 16.2 $\pm $ 0.1   \\
$^{114}$Sn    & 15.03      & 15.20   & 14.70     & 16.1 $\pm $ 0.1   \\
$^{116}$Sn    & 14.94      & 15.08   & 14.56     & 15.8 $\pm $ 0.1   \\
$^{118}$Sn    & 14.82      & 15.13   & 14.48     & 15.8 $\pm $ 0.1   \\
$^{120}$Sn    & 14.69      & 15.08   & 14.58     & 15.7 $\pm $ 0.1   \\
$^{122}$Sn    & 14.68      & 15.00   & 14.61     & 15.4 $\pm $ 0.1   \\
$^{124}$Sn    & 14.68      & 14.96   & 14.51     & 15.3 $\pm $ 0.1   \\
\bottomrule
\end{tabularx}
\end{table}

\vspace{-10pt}
\begin{table}[H]
\caption{The values of the centroid energies $E_{ISGMR}$ (in MeV)
of Pb isotopes obtained from HF+CDFM calculations in this work
using SLy4 and SkM Skyrme forces compared with the experimental
data found in the literature.\label{tab3}}
\newcolumntype{C}{>{\centering\arraybackslash}X}
\begin{tabularx}{\textwidth}{CCCCC}
\toprule
\textbf{Nucleus} & \textbf{SLy4}  & \textbf{SkM}  & \textbf{Exp.}  & \textbf{Theory}         \\
\midrule
$^{204}$Pb        & 12.16      & 12.29   & 13.98 \cite{Fujiwara2011}               &          \\
$^{206}$Pb        & 12.12      & 12.23   & 13.94 \cite{Fujiwara2011}               &          \\
$^{208}$Pb        & 12.10      & 12.15   & 13.96 $\pm $ 0.2 \cite{Youngblood2004}  & 14.453 \cite{Chen2012}  \\
\bottomrule
\end{tabularx}
\end{table}

\begin{figure}[H]
\includegraphics[width=10.5cm]{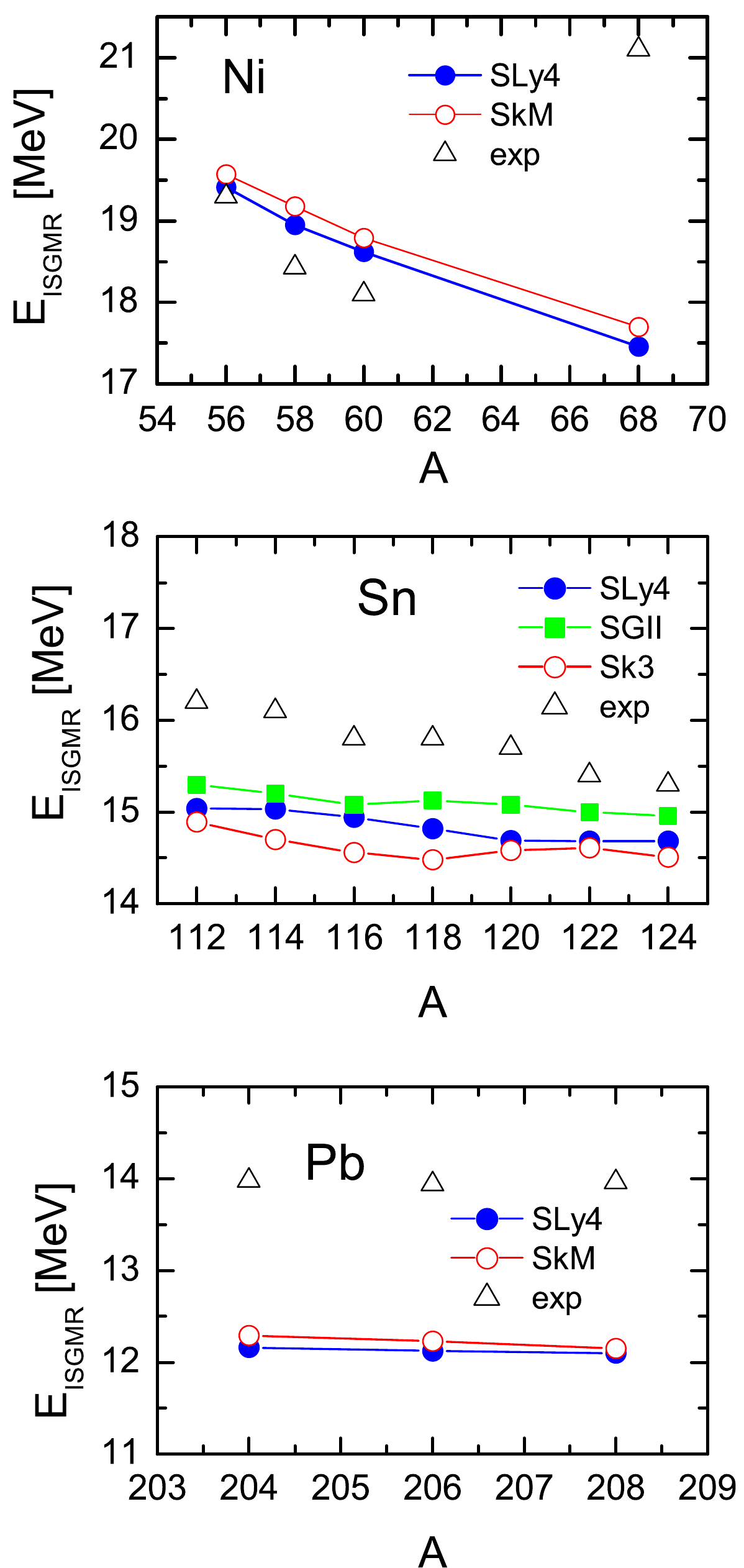}
\caption{The centroid energies $E_{ISGMR}$ as a function of the
mass number $A$ for Ni, Sn, and Pb isotopes in the cases of SLy4,
SGII, Sk3, and SkM forces and $r_{0}=1.2$ fm (Equation~(\ref{eq:1b}))
compared with the experimental data.
\label{fig2}}
\end{figure}

The collective (bulk) character of the giant resonances and
nuclear incompressibility presumes a quite smooth variation of the
properties of the ISGMR with mass, thus not expecting very strong
variations related to the internal nuclear structure. The isotopic
evolution of the centroid energies $E_{ISGMR}$ for the Ni, Sn, and
Pb isotopes is presented in Figure~\ref{fig2} in the case when $r_{0}=1.2$
fm is used. In general, as expected, a smooth decrease in the excitation energies
of the ISGMR with the increase in the mass number $A$ is observed for the
three isotopic chains and for all Skyrme forces used in the
calculations. Furthermore, going from Ni to Pb isotopic chain the ``gap''
between our results and the corresponding experimental data
becomes larger in a way that the obtained values of $E_{ISGMR}$
underestimate the experimentally extracted values. Nevertheless,
this difference does not exceed 1--2 MeV in the case of Sn and Pb
isotopes and practically is minimal for Ni isotopes.

\begin{figure}[H]
\includegraphics[width=10cm]{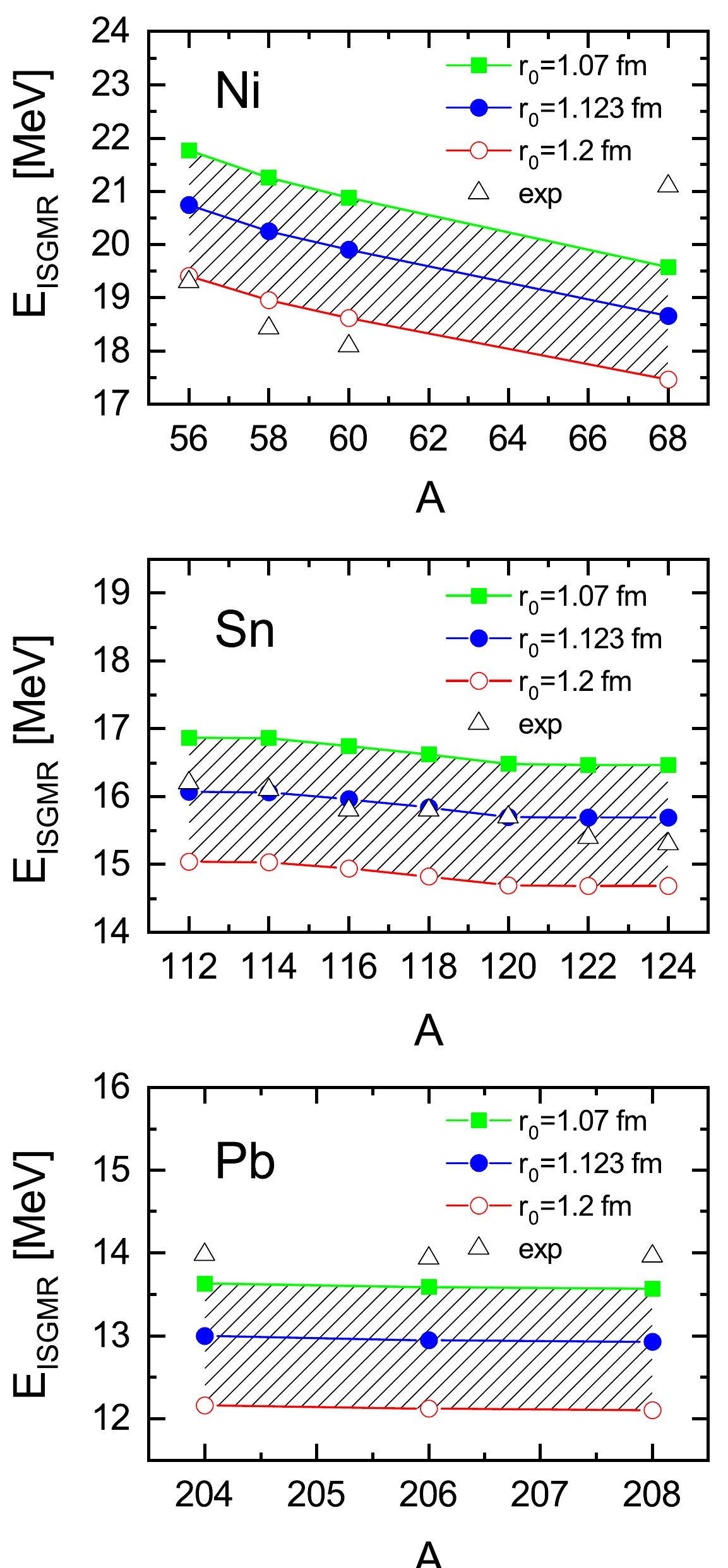}
\caption{The centroid energies $E_{ISGMR}$ as a function of the
mass number $A$ for Ni, Sn, and Pb isotopes in the case of SLy4
force obtained with three different values of the parameter
$r_{0}=1.07, 1.123, 1.2$ fm (Equation~(\ref{eq:1b})) compared with the
experimental data.
\label{fig3}}
\end{figure}

As a test of the role of the half-density radius parameter $r_{0}$
on the centroid energy (Equation~(\ref{eq:1b})), we present in
Figure~\ref{fig3} the results of $E_{ISGMR}$ for the same Ni, Sn,
and Pb isotopic chains in the case of SLy4 force obtained with two
more values of $r_{0}$. In addition to the results with
$r_{0}=1.2$ fm (e.g., in Refs.~\cite{Brown84,Li2009}) given in
Figure~\ref{fig2}, the values of $E_{ISGMR}$ calculated with
$r_{0}=1.07$ fm (for instance, in Ref.~\cite{Walecka95}) and
$r_{0}=1.123$ fm \cite{Eisenberg70} are shown in Figure~\ref{fig3}.
It is seen from the figure that with the increase of $r_{0}$
the agreement with the experimental data becomes better for
lighter isotopes. Particularly, the value of \linebreak $r_{0}=1.123$ fm
leads to fair agreement of the ISGMR energies for Sn isotopes,
while for Ni isotopes the experimental data are reproduced better
with $r_{0}=1.2$ fm and for Pb isotopes with $r_{0}=1.07$ fm. Here
we would like to note that the specific choice of the $r_{0}$
parameter values adopted to calculate the values of the centroid
energies by using expression (\ref{eq:1b}) is often used in the
literature. The values of the measured nuclear radii are
deduced from processes with strongly interacting particles or
electron (muon) scattering. It is well known that the $A$-dependence
of $r_{0}$ exhibits a smooth decrease with $A$ being 1.07 fm for nuclei with $A>16$
and increasing to 1.2 fm for heavy nuclei. This results on
the calculated values of $E_{ISGMR}$ and the corresponding ranges of change
in respect to $r_{0}$ are illustrated in Figure~\ref{fig3} by hatched areas. Thus,
we find a sensitivity of the results for centroid energies of ISGMR to the
radial parameter $r_{0}$ and this fact has to be taken into
account when considering resonances in light, medium, and heavy
nuclei.

\unskip

%%%%%%%%%%%%%%%%%%%%%%%%%%%%%%%%%%%%%%%%%%
\section{Conclusions and Perspectives}\label{sec:4}

We have performed a systematic study of the isoscalar giant
monopole resonance in Ni, Sn, and Pb isotopes within the
microscopic self-consistent Skyrme HF+BCS method and coherent
density fluctuation model. In the present calculations four
different Skyrme parameter sets are used: SLy4, SGII, Sk3, and
SkM. They are chosen since they were employed in our previous
works and, more importantly, are characterized by different values
of the nuclear matter incompressibility. The calculations are
based on the Brueckner energy-density functional for nuclear
matter.

A very good agreement is achieved between the calculated centroid
energies of the ISGMR and corresponding experimental values for Ni
isotopes when $r_{0}=1.2$ fm. Especially this concerns the exotic double-magic
$^{56}$Ni nucleus, for which the obtained (with SLy4 Skyrme force)
value is 19.41 MeV, in consistency with the centroid position of
the ISGMR found at $19.1\pm 0.5$ MeV. For $^{68}$Ni our
predictions for $E_{ISGMR}$ with both Skyrme interactions are
rather below the experimental result, obviously requiring a larger
value of $\Delta K$. The comparative analysis of the centroid
energies in the case of Sn and Pb isotopes shows less agreement with
$r_{0}=1.2$ fm, but still in an acceptable limits. This could be partly
due to the chosen physical criterion that is applied to calculate the finite
nucleus incompressibility (Equation~(\ref{eq:18})). The latter point
will be a subject of future study. The agreement with the
experimental values of $E_{ISGMR}$ can be improved also by varying the
parameter $r_{0}$ (Equation~(\ref{eq:1b})) in strong connection with
the mass dependence of this parameter and its effect for the considered
isotopes.

In general, the results obtained in the present work demonstrate
the relevance of our theoretical approach to probe the excitation
energy of the ISGMR in various nuclei. Our future goal is to
extend this theoretical study by employing more realistic
energy-density functionals for nuclear matter, from one side. For
example, the role of microscopic three-body forces in the proposed
approach to study the giant monopole resonances can be clearly
revealed by applying the latest version of the
Barcelona--Catania--Paris--Madrid nuclear EDF (\cite{Sharma2015} and
references therein) and particularly to treat successfully medium-heavy
nuclei. In addition, a good choice could be the microscopic EOS
derived by Sammarruca {et al.} \cite{Sammarruca2015} based on
high-precision chiral nucleon-nucleon potentials at
next-to-next-to-next-to-leading order (N$^{3}$LO) of chiral
perturbation theory \cite{Machleidt2011,Entem2003}. Thus, by
employing of microscopic input in the energy-density functionals
for nuclear matter, a stronger connection with fundamental nuclear
forces can be achieved. From another side, the important issue
will be to expand the nuclear spectrum to lighter and medium mass
nuclei considering also deformed nuclei, in which the breaking of
spherical symmetry would play a role. In addition, to extract the
isospin dependence of the incompressibility coefficient, a key
ingredient in astrophysical studies, further theoretical
investigations are needed to carry out calculations of the ISGMR
for neutron-rich nuclei and to compare the results with the
available experimental data.

%This section is not mandatory, but can be added to the manuscript
%if the discussion is unusually long or complex.
\vspace{+6pt}
%%%%%%%%%%%%%%%%%%%%%%%%%%%%%%%%%%%%%%%%%%
\authorcontributions{Conceptualization, M.K.G., M.V.I. and A.N.A.; Formal analysis,
M.K.G., M.V.I. and A.N.A.; Visualization, M.K.G. and M.V.I.;
Writing-original draft, M.K.G.; Writing-review and editing,
M.K.G., M.V.I., Y.I.K. and A.N.A. All authors have read and agreed
to the published version of the manuscript.}
%MDPI: For research articles with several authors, a short paragraph specifying their individual contributions must be provided. The following statements should be used
%``Conceptualization, X.X. and Y.Y.; methodology, X.X.; software, X.X.; validation, X.X., Y.Y. and Z.Z.; formal analysis, X.X.; investigation, X.X.; resources, X.X.; data %curation, X.X.;
%writing---original draft preparation, X.X.; writing---review and editing, X.X.; visualization, X.X.; supervision, X.X.; project administration, X.X.; funding acquisition, Y.Y.
%All authors have read and agreed to the published version of the
%manuscript.'', please turn to the
%\href{http://img.mdpi.org/data/contributor-role-instruction.pdf}{CRediT
%taxonomy} for the term explanation. Authorship must be limited to
%those who have contributed substantially to the work~reported.

\funding{This research was funded by the Bulgarian National Science Fund under Contract No.~KP-06-N38/1.}

\institutionalreview{Not applicable.}
%MDPI: In this section, you should add the Institutional Review Board Statement and approval number, if relevant to your study. You might choose to exclude this statement if the study did not require ethical approval. Please note that the Editorial Office might ask you for further information. Please add “The study was conducted in accordance with the Declaration of Helsinki, and approved by the Institutional Review Board (or Ethics Committee) of NAME OF INSTITUTE (protocol code XXX and date of approval).” for studies involving humans. OR “The animal study protocol was approved by the Institutional Review Board (or Ethics Committee) of NAME OF INSTITUTE (protocol code XXX and date of approval).” for studies involving animals. OR “Ethical review and approval were waived for this study due to REASON (please provide a detailed justification).” OR “Not applicable” for studies not involving humans or animals.

\informedconsent{Not applicable.}%MDPI: Any research article describing a study involving
%humans should contain this statement. Please add ``Informed
%consent was obtained from all subjects involved in the study.'' OR
%``Patient consent was waived due to REASON (please provide a
%detailed justification).'' OR ``Not applicable'' for studies not
%involving humans. You might also choose to exclude this statement
%if the study did not involve humans. Written informed consent for publication must be obtained from participating patients who can be identified (including by the patients themselves). Please state ``Written informed consent has been obtained from the patient(s) to publish this paper'' if applicable.

\dataavailability{Not applicable.}%MDPI: In this section, please provide details regarding where data supporting reported results can be found, including links to publicly archived datasets analyzed or generated during the study. Please refer to suggested Data Availability Statements in section ``MDPI Research Data Policies'' at \url{https://www.mdpi.com/ethics}. If the study did not report any data, you might add ``Not applicable'' here.

%\acknowledgments{M.K.G., M.V.I., and A.N.A are grateful for the
%support of the Bulgarian National Science Fund under Contract
%No.~KP-06-N38/1.}

\conflictsofinterest{The authors declare no conflict of interest.}%MDPI: Declare conflicts of interest or state ``The authors declare no conflict of interest.'' Authors must identify and declare any personal circumstances or interest that may be perceived as inappropriately influencing the representation or interpretation of reported research results. Any role of the funders in the design of the study; in the collection, analyses or interpretation of data; in the writing of the manuscript, or in the decision to publish the results must be declared in this section. If there is no role, please state ``The funders had no role in the design of the study; in the collection, analyses, or interpretation of data; in the writing of the manuscript, or in the decision to publish the~results''.

%% Optional
%\sampleavailability{Samples of the compounds ... are available from the authors.}

%%%%%%%%%%%%%%%%%%%%%%%%%%%%%%%%%%%%%%%%%%
%% Only for journal Encyclopedia
%\entrylink{The Link to this entry published on the encyclopedia platform.}

%%%%%%%%%%%%%%%%%%%%%%%%%%%%%%%%%%%%%%%%%%
%% Optional
%\abbreviations{Abbreviations}{
%The following abbreviations are used in this manuscript:\\

%\noindent
%\begin{tabular}{@{}ll}
%MDPI & Multidisciplinary Digital Publishing Institute\\
%DOAJ & Directory of open access journals\\
%TLA & Three letter acronym\\
%LD & Linear dichroism
%\end{tabular}}

%%%%%%%%%%%%%%%%%%%%%%%%%%%%%%%%%%%%%%%%%%
\begin{adjustwidth}{-\extralength}{0cm}
%\printendnotes[custom] % Un-comment to print a list of endnotes

\reftitle{References}%MDPI: Important Note: Please confirm and add all titles for journal type bwlow like reference 3.

\PublishersNote{}
\end{adjustwidth}
\end{document}